\documentclass{article}
\usepackage{epsf}
\usepackage{times}
\usepackage{amsmath}
\usepackage{amssymb}

\newcommand{\braket}[2]{\langle #1 | #2 \rangle}
\newcommand{\Tr}{{\mathrm{Tr}}}
\newcommand{\rank}{{\mathrm{rank}}}
\newcommand{\ket}[1]{\left | \, #1 \right \rangle}
\newcommand{\smallket}[1]{ | \, #1 \rangle}
\newcommand{\bra}[1]{\left \langle #1 \, \right |}
\newcommand{\smallbra}[1]{\langle #1 \, |}
\newcommand{\proj}[1]{| {#1}\rangle \! \langle{#1}|}
\newcommand{\qed}{\hfill $\square$}

\newtheorem{thm}{Theorem}

\newtheorem{lemma}[thm]{Lemma}

\begin{document}
\begin{center}
{\Large Equivalence of Additivity Questions \\ in Quantum
Information Theory}\\
{\large Peter W. Shor}\\
AT\&T Labs Research\\
Florham Park, NJ 07922 USA
\end{center}

\begin{center}
Abstract
\end{center}
{\em We reduce the number of open additivity problems in quantum
information theory by showing that four of them are equivalent.
Namely, we show that the conjectures of
additivity of the minimum output entropy of a quantum channel,
additivity of the Holevo expression for the
classical capacity of a quantum channel, 
additivity of the entanglement of formation, 
and strong superadditivity of the entanglement of formation,
are either all true or all false.}

\section{Introduction}
\label{sec-intro}

The study of quantum information theory has led to
a number of seemingly related open questions 
that center around whether certain quantities
are additive.  We show that four of these questions are equivalent.
In particular, we show that the four conjectures of 
\begin{enumerate}
\item[i.] additivity of the minimum entropy output of a quantum channel, 
\item[ii.] additivity of the Holevo capacity of a quantum channel, 
\item[iii.] additivity of the entanglement of formation,
\item[iv.] strong superadditivity of the entanglement of formation, 
\end{enumerate}
are either all true or all false.  

Two of the basic ingredients in our proofs are already known.  The first
is an observation of Matsumoto, Shimono and Winter \cite{MSW} that the
Stinespring dilation theorem relates a constrained version of the 
Holevo capacity formula to the entanglement of formation.  
The second is the realization that the entanglement of formation (or the
constrained Holevo capacity) is a linear programming problem, and so 
there is also a dual linear formulation.  
This formulation was first presented by Audenaert and Braunstein \cite{AB}, 
who expressed it in the language of convexity rather than that of linear 
programming.
We noted this independently \cite{Shor-compute}.
These two ingredients are explained in Sections \ref{sec-MSW} 
and \ref{sec-lp}.  

The rest of this paper is organized as follows.  Section~\ref{sec-background}
gives some background in quantum information theory,
describes the additivity questions we consider, and gives brief histories
of them.  Sections~\ref{sec-MSW} and \ref{sec-lp} explain the two 
ingredients we describe above, and are positioned immediately before
the first sections in which they are used.
To show that the conditions (i) to (iv) are equivalent, 
in Section \ref{sec-ii-iii} we prove that (ii) $\rightarrow$ (iii):
additivity of 
the Holevo capacity implies additivity of entanglement of formation.
In Section \ref{sec-iii-iv} we prove (iii) $\rightarrow$ (iv): 
additivity of entanglement of formation implies strong superadditivity 
of entanglement of formation.  This implication was independently discovered
by Pomeransky \cite{ssa-from-aef}.  In Section \ref{sec-i-iii} we prove that 
(i) $\rightarrow$ (iii): additivity of minimum entropy output implies 
additivity of entanglement of formation.  In Section \ref{sec-iv-all}, 
we give simple proofs showing that (iv) $\rightarrow$ (i), 
(iv) $\rightarrow$ (ii), and (iv) $\rightarrow$ (iii).
The first implication is the only one that was not in the literature, 
and we assume this is mainly because nobody had tried to prove it.  
The second
of these implications was already known, but for completeness we give 
a proof. 
The third of these implications is trivial.\footnote{In fact, 
property (iv), strong superadditivity of $E_F$, seems to be in some 
sense the ``strongest'' of these equivalent statements, as 
it is fairly easy to show that strong superadditivity of entanglement 
of formation implies the other three additivity results 
whereas the reverse directions appear to require substantial work.
Similarly, property (i) appears to be the ``weakest'' of these
statements.}  
In Section \ref{sec-all-i} we give
proofs that (ii) $\rightarrow$ (i) and (iii) $\rightarrow$ (i): 
either additivity of the Holevo capacity or of 
the entanglement of formation
implies additivity of the minimum entropy output.  
These implications
complete the proof of equivalence.  Strictly speaking, the only 
implications we need for the proof of equivalence are those in 
Sections~\ref{sec-iii-iv}--\ref{sec-all-i}.  
We include the proof in Section~\ref{sec-ii-iii} 
because it uses one of the techniques used later for 
Section~\ref{sec-i-iii} without introducing the extra complexity of
the dual linear programming formulation.  Finally, in 
Section \ref{sec-discuss} we comment on the implications of the 
results in our paper and give some open problems.  

\section{Background and Results}
\label{sec-background}

One of the important intellectual breakthroughs of the 20th century
was the discovery and development of information theory.  A cornerstone 
of this field is Shannon's proof that a communication
channel has a well-defined information carrying capacity and his
formula for calculating it.  For communication channels
that intrinsically incorporate quantum effects, this classical theory is no 
longer valid.  The search for the proof of the analogous quantum formulae 
is a subarea of quantum information theory that has recently received
much study.

In the generalization of Shannon theory to the quantum
realm, the definition of a stochastic communication channel generalizes 
to a completely positive trace-preserving linear map (CPT map).
We call such a map a {\em quantum channel}.  
In this paper, we consider only finite-dimensional CPT maps; these take
$d_{\mathrm{in}} \times d_{\mathrm{in}}$ Hermitian matrices to
$d_{\mathrm{out}} \times d_\mathrm{out}$ Hermitian 
matrices.  In particular, these maps take density matrices
(trace 1 positive semidefinite matrices) to density matrices.
Note that the input dimension can be different from
the output dimension, and that these dimensions are both finite.  
Infinite dimensional quantum channels (CPT maps) are both important 
and interesting, but dealing with them also introduces extra complications
that are beyond the scope of this paper.

There are several characterizations of CPT maps.  We need the 
characterization given 
by the Stinespring dilation theorem, which says that every CPT map can
be described by an unitary embedding followed by a partial trace.  
In particular,
given a finite-dimensional CPT map $N$, we can express it as
\[
N(\rho) = \Tr_B U(\rho)
\]
where $U(\rho)$ is a unitary embedding, i.e., there is some ancillary
space ${\cal H}_B$ such that $U$ takes 
${\cal H}_{\mathrm{in}}$ to 
${\cal H}_\mathrm{out} \otimes {\cal H}_B$ by
\[
U(\rho) = V \rho V^\dag
\]
and $V$ is a unitary matrix mapping ${\cal H}_\mathrm{in}$ to
$\mathrm{range}(V) \subseteq {\cal H}_\mathrm{out} \otimes {\cal H}_B$.
We also need the operator sum characterization of CPT maps.  This
characterization says that any finite-dimensional CPT map $N$ can be 
represented as
\[
N(\rho) = \sum_k A_k \rho A_k^\dag
\]
where the $A_k$ are complex matrices satisfying
\[
\sum A_k^\dag A_k = I .
\]

The Holevo information\footnote{This has also been called the
Holevo bound and the Holevo $\chi$-quantity.}
$\chi$ is a quantity which is associated with a probabilistic
ensemble of quantum states (density matrices).  
If density matrix $\rho_i$ occurs in the ensemble with probability $q_i$, 
the Holevo information $\chi$ of the ensemble is
\[
\chi = H(\sum_i q_i \rho_i) - \sum_i q_i H(\rho_i)
\]
where $H$ is the von Neumann entropy $H(\rho) = - \Tr\, \rho \log \rho$.
This quantity was introduced in \cite{Gordon,Levitin,Hol73}
as a bound for the amount of information extractable by measurements
from this ensemble of 
quantum states.  The first published proof of this bound was given by
Holevo \cite{Hol73}.  It was much later shown that maximizing
the Holevo capacity over all probabilistic
ensembles of a set of quantum states gives the
information transmission
capacity of this set of quantum states; more specifically, this is
the amount of classical information which can be transmitted 
asymptotically per quantum state by using codewords that are
tensor products of these quantum states,
as the length of these codewords goes to infinity 
\cite{H,SW}.  Optimizing $\chi$
over ensembles composed of states that are potential outputs of a 
quantum channel 
gives the quantum capacity of this quantum channel over a restricted set 
of protocols, namely those protocols which are not allowed to send inputs
entangled between different channel uses.  If the channel is $N$, we
call this quantity $\chi_N$; it is defined as
\begin{equation}
\chi_N =
\max_{\{p_i,\ket{v_i}\}}  
H(N(\sum_i p_i \proj{v_i})) - \sum_i p_i H(N(\proj{v_i})),
\label{def-chi}
\end{equation}
where the maximization is over ensembles $\{p_i, \, \ket{v_i} \}$ where
$\sum_i p_i = 1$ and $\ket{v_i} \in {\cal H}_\mathrm{in}$, the input
space of the channel $N$.   

The regularized Holevo capacity is
\[
\lim_{n \rightarrow \infty} \frac{1}{n} \chi_{N^{\otimes n}};
\]
this gives the capacity of a quantum channel to transmit classical 
information when inputs entangled between different channel uses are
allowed.  The question of whether
the quantum capacity is given by the single-symbol Holevo capacity $\chi_N$
is the question of whether the capacity $\chi_N$ is additive; that is, whether
\[
\chi_{N_1 \otimes N_2} = \chi_{N_1} + \chi_{N_2}.  
\]
The $\geq$ relation is easy; the open question is the $\leq$ relation.

The question of additivity of the minimum entropy output of a quantum 
channel was originally considered independently by several people, 
including the author, and appears to have been first considered in print in
\cite{King-Ruskai}.  It was originally posed as a possible first step to
proving additivity of the Holevo capacity $\chi_N$. 
The question is whether
\[
\min_{\ket{\phi}} H(N_1 \otimes N_2(\proj{\phi})) =
\min_{\ket{\phi}} H(N_1(\proj{\phi})) + \min_{\ket{\phi}} H(N_2(\proj{\phi})),
\]
where the minimization ranges over states $\ket{\phi}$ in the
input space of the channel.  Note that by the concavity of the von Neumann 
entropy, if we minimize over
mixed states $\rho$---i.e., $\min_\rho H(N(\rho))$---there will always be 
a rank one $\rho = \proj{\phi}$ achieving the minimum.

The statements (iii) and (iv) in our equivalence theorem both deal with
entanglement. 
This is one of the stranger phenomena of quantum mechanics.  
Entanglement occurs when two (or more) quantum systems are non-classically
correlated.  The canonical example of this phenomenon is an EPR pair.
This is the state of two quantum systems (called qubits, as they are each
two-dimensional):
\[
\frac{1}{\sqrt{2}} \big( \ket{01} - \ket{10} \big).
\]
Measurements on each of these two qubits separately can exhibit 
correlations which cannot be modeled by two separated classical 
systems \cite{Bell}.  

A topic in quantum information theory that has recently attracted 
much study is that of quantifying entanglement.
The entanglement of a bipartite pure state is easy to define and compute; 
this is the entropy 
of the partial trace over one of the two parts 
\[
E_\mathrm{pure}(\proj{v}) = H(\Tr_B \proj{v}).
\]
Asymptotically, two parties sharing $n$ copies of a bipartite
pure state $\proj{v}$ can use local quantum operations and classical
communication (called {\em LOCC operations}) to produce 
$n E_\mathrm{pure}(\proj{v}) - o(n)$
nearly perfect EPR pairs, and can similarly form $n$ nearly perfect
copies of $\proj{v}$ from $n E_\mathrm{pure}(\proj{v}) + o(n)$ EPR pairs
\cite{ent-conc}.  This implies that a for pure state $\proj{v}$, 
the entropy of the partial trace is the natural quantitative measure of 
the amount of entanglement contained in $\proj{v}$.

For mixed states (density matrices of rank $>$ $1$), 
things become more 
complicated.  The amount of pure state entanglement asymptotically 
extractable from a state using LOCC operations (the {\em distillable 
entanglement}) is now no longer necessarily equal to the amount of 
pure state entanglement asymptotically
required to create a state using LOCC operations (the {\em entanglement cost}) 
\cite{VDC}.  In general, the entanglement cost must be at least
the distillable entanglement, as LOCC operations cannot increase the
amount of entanglement.  

The {\em entanglement of formation} was introduced in \cite{BDSV}. 
Suppose we have a bipartite state $\sigma$ on a
Hilbert space ${\cal H}_A \otimes {\cal H}_B$.  The entanglement of 
formation is
\begin{equation}
E_F(\sigma) = \min_{\{p_i,\, \ket{v_i}\}} \sum_i p_i H(\Tr_B \proj{v_i})
\label{def-ef}
\end{equation}
where the minimization is over all ensembles such that 
$\sum_i p_i \proj{v_i} = \sigma$ with probabilities $p_i$ satisfying
$\sum_i p_i = 1$.
The entanglement of formation must be at least the entanglement cost,
as the decomposition of the state $\sigma$ yielding $E_F(\sigma)$ can be 
used to create a prescription for asymptotically constructing 
$\sigma^{\otimes n}$ from $n E_F(\sigma) + o(n)$ EPR pairs.  
The regularized entanglement of formation
\[
\lim_{n \rightarrow \infty} \frac{1}{n} E_F(\sigma^{\otimes n})
\]
has been proven to give the entanglement cost of 
a quantum state \cite{HHT}.  
As in the case of channel capacity,
a proof of additivity, i.e., that
\[
E_F(\sigma_1 \otimes \sigma_2) =
E_F(\sigma_1) + E_F(\sigma_2),
\]
would imply that regularization is not necessary.

The question of strong superadditivity of entanglement of formation has been 
previously
considered in \cite{BenNarn,VDC,MSW,AB}.  This conjecture says that
for all states $\sigma$ over a quadripartite 
system ${\cal H}_{A1} \otimes {\cal H}_{A2} \otimes {\cal H}_{B1} \otimes {\cal H}_{B2}$, we have
\[
E_F(\sigma) \geq E_F(\Tr_2 \sigma)  + E_F(\Tr_1 \sigma)
\]
where the entanglement of formation $E_F$ is taken over the bipartite A-B
division, as in (\ref{def-ef}).  
This question was originally considered in relation to the 
question of additivity of $E_F$.  
The strong superadditivity of entanglement
of formation is known to imply both the additivity of entanglement
of formation (trivially) and the additivity of Holevo capacity of a 
channel \cite{MSW}.
A proof similar to ours that additivity 
of $E_F$ implies strong superadditivity of $E_F$ was discovered
independently; it appears in \cite{ssa-from-aef}.

We can now state the main result of our paper.
\begin{thm}
The following are equivalent.
\begin{enumerate}
\item[i.] The additivity of the minimum entropy output of a quantum channel.
Suppose we have two quantum channels (CPT maps) $N_1$ (taking 
$\mathbb{C}^{\,d_{1,\mathrm{in}} \times d_{1,\mathrm{in}} }$ to
$\mathbb{C}^{\,d_{1,\mathrm{out}} \times d_{1,\mathrm{out}} }$) and $N_2$
(taking
$\mathbb{C}^{\,d_{2,\mathrm{in}} \times d_{2,\mathrm{in}} }$ to
$\mathbb{C}^{\,d_{2,\mathrm{out}} \times d_{2,\mathrm{out}} }$).
Then
\[
\min_{{\ket{\phi}}} H((N_1\otimes N_2)({\proj{\phi}})) =
\min_{{\ket{\phi}}} H(N_1(\proj{\phi})) +  
\min_{{\ket{\phi}}} H(N_2({\proj{\phi}}))
\]
where $H$ is the von Neumann entropy and the minimization is taken over
all vectors $\ket{\phi}$ in the input space of the channels.  
\item[ii.]   The additivity of the Holevo capacity of a quantum channel, 
Assume we have two quantum channels $N_1$ and $N_2$, as in (i).  Then
\[
\chi_{N_1\otimes N_2} =
\chi_{N_1} +  \chi_{N_2},
\]
where $\chi$ is defined as in Eq.~(\ref{def-chi}).
\item[iii.] Additivity of the entanglement of formation.
Suppose we have two quantum states 
$\sigma_1 \in {\cal H}_{A1} \otimes {\cal H}_{B1}$
and $\sigma_2 \in {\cal H}_{A2} \otimes {\cal H}_{B2}$.
Then
\[
E_F(\sigma_1 \otimes \sigma_2) = E_F(\sigma_1) + E_F(\sigma_2),
\]
where $E_F$ is defined as in Eq.~(\ref{def-ef}).  In particular,
the entanglement of formation is calculated over the bipartite $A$--$B$ 
partition.
\item[iv.] The strong superadditivity of the entanglement of formation.
Suppose we have a density matrix $\sigma$ over a quadripartite system
system 
${\cal H}_{A1} \otimes {\cal H}_{A2} \otimes {\cal H}_{B1} 
\otimes {\cal H}_{B2}$.  Then
\[
E_F(\sigma) \geq E_F(\Tr_2 \sigma) + E_F(\Tr_1 \sigma),
\]
where the entanglement of formation is calculated over the 
bipartite $A$--$B$ partition.  Here, the operator 
$\Tr_1$ traces out the space ${\cal H}_{A1}\otimes {\cal H}_{B1}$, and
$\Tr_2$ traces out the space ${\cal H}_{A2}\otimes {\cal H}_{B2}$.
\end{enumerate}
\end{thm}

\section{The correspondence of Matsumoto, Shimono and Winter}
\label{sec-MSW}

Recall the definition of the Holevo capacity for a channel $N$:
\[
\chi_N = \max_{\{p_i,\, \ket{\phi_i}\}} 
H(N( \sum_i p_i \proj{\phi_i})) -
\sum_i p_i H(N( \proj{\phi_i}))
\]
Recall also the definition of entanglement of formation.  For
a bipartite state $\sigma$ on ${\cal H}_A \otimes {\cal H}_B$, the
entanglement of formation is
\[
E_F(\sigma) = \min_{\{{p_i,\, \ket{v_i}\}} \atop {\sum p_i \proj{v_i} = \sigma}}
\sum_i p_i H(\Tr_B \proj{v_i})
\]

Let us define a constrained version of the Holevo capacity, which is just the
Holevo capacity over ensembles whose average input is $\rho$.
\begin{equation}
\chi_N(\rho) = \max_{\{{p_i,\, \ket\phi_i\}}\atop {\sum_i p_i 
\proj{\phi_i} = \rho}} H(N( \sum_i p_i \proj{\phi_i})) -
\sum_i p_i H(N( \proj{\phi_i})) 
\label{constrainedHolevocap}
\end{equation}

The paper of Matsumoto, Shimono and Winter \cite{MSW} gives a connection 
between this constrained version of the Holevo capacity and the 
entanglement of formation, which we now explain.  The Stinespring dilation
theorem says that any quantum channel can be realized as a unitary 
transformation followed by a partial trace.   Suppose we have 
a channel $N$ taking ${\cal H}_\mathrm{in}$ to ${\cal H}_A$.  We can
find a unitary embedding $U(\rho) = V \rho V^\dag$ that takes 
${\cal H}_\mathrm{in}$ to 
${\cal H}_A \otimes {\cal H}_B$ such that
\[
N(\mu) = \Tr_B U(\mu)
\]
for all density matrices $\mu \in {\cal H}_\mathrm{in}$.
Now, $U$ maps an ensemble of input states $\{ p_i, \ket{\phi_i}\}$ with
$\rho = \sum_i p_i \proj{\phi_i}$ to an ensemble of states 
$\{ p_i, \, \ket{v_i}  = V \ket{\phi_i} \}$ 
on the bipartite system ${\cal H}_A \otimes {\cal H}_B$ such that
$\sum_i p_i \proj{v_i} = \sigma = U(\rho)$.  

Conversely, if we are given a bipartite state 
$\sigma \in {\cal H}_A \otimes {\cal H}_B$, we can find an input space
${\cal H}_\mathrm{in}$ with $\dim {\cal H}_\mathrm{in} = \rank \, \sigma$,
a density matrix $\rho \in {\cal H}_\mathrm{in}$, and a unitary 
embedding $U: {\cal H}_{\mathrm{in}}\rightarrow{\cal H}_{\mathrm{out}}$ 
such that
$U(\rho) = \sigma$.  We can then define $N$ by
\[
N(\mu) = \Tr_B U(\mu),
\]
establishing the same relation between $N$, $U$, $\rho$ and $\sigma$. 
Note that since we chose  
$\dim {\cal H}_\mathrm{in} = \rank\, \sigma = \rank\, \rho$,
$\rho$ has full rank in ${\cal H}_\mathrm{in}$.

Since
$N(\proj{\phi_i}) = \Tr_B \proj{v_i}$, we have
\[
\chi_N(\rho) = H(N(\rho)) - E_F(\sigma).
\]
Now, suppose $E_F(\sigma)$ is additive.  I claim that
$\chi_N(\rho)$ is as well, and vice versa.  Let us take 
$N_1(\rho) = \Tr_B U_1(\rho) $ and 
$N_1(\rho) = \Tr_B U_2(\rho) $.  If $U_1(\rho_1) = \sigma_1$ and 
$U_2(\rho_2) = \sigma_2$, then we have
\begin{eqnarray*}
\chi_{N_1 \otimes N_2}(\rho_1 \otimes \rho_2) &=&
H(N_1 \otimes N_2(\rho_1 \otimes \rho_2)) - E_F(\sigma_1 \otimes \sigma_2)\\
&=& H(N_1(\rho_1)) + H(N_2(\rho_2)) - E_F(\sigma_1 \otimes \sigma_2)\\
\end{eqnarray*}
The first term on the right-hand side is additive, so the entanglement
of formation $E_F$ is additive
if and only if the constrained capacity $\chi_N(\rho)$ is.

\section{Additivity of $\chi$ implies additivity of $E_F$}
\label{sec-ii-iii}

Recall the definition of the Holevo capacity for a channel $N$:
\[
\chi_N = \max_{\{p_i,\, \ket{\phi_i}\}} 
H(N( \sum_i p_i \proj{\phi_i})) -
\sum_i p_i H(N( \proj{\phi_i}))
\]
where the maximization is over ensembles $\{p_i,\ket{\phi_i}\}$
with $\sum_i p_i = 1$.
Recall also our definition of
a constrained version of the Holevo capacity, which is just the 
definition of
the Holevo capacity with the maximization only over ensembles whose 
average input is $\rho$.
\[
\chi_N(\rho) = \max_{{\{p_i,\, \ket{\phi_i}\}}\atop {\sum_i p_i 
\proj{\phi_i} = \rho}} H(N( \sum_i p_i \proj{\phi_i})) -
\sum_i p_i H(N( \proj{\phi_i})) 
\]

Let $\sigma$ be the state whose entanglement of formation we are
trying to compute.  The MSW correspondence yields
a channel $N$ and an input state $\rho$ so that
\[
N(\rho) = \Tr_B \sigma
\]
and 
\[
\chi_N(\rho) = H(N(\rho)) - E_F(\sigma)
\]
This is very nearly the channel capacity, the only difference being
that the $\rho$ above is not necessarily the $\rho$ that maximizes
$\chi_N$.  
Only one element is missing for the proof that additivity of channel 
capacity implies additivity of entanglement of formation:
namely making sure that the average density matrix for the ensemble 
giving the optimum channel capacity is equal to a desired matrix $\rho_0$.  
This cannot be done directly \cite{Ruskai}, but we solve the problem 
indirectly.

We now give the intuition for our proof.  Suppose we could 
define a new channel $N'$ which,
instead of having capacity
\[
\chi_N = 
\max_{\rho} \chi_N(\rho)
\]
has capacity 
\begin{equation}
\chi_{N'} = 
\max_{\rho} \chi_N(\rho) + \Tr\, \rho \tau
\label{whatwewant}
\end{equation}
for some fixed Hermitian matrix $\tau$.  For a proper choice of $\tau$, this
will ensure that the maximum of this channel occurs at the desired
$\rho$.  Consider two entangled states $\sigma_1$ and 
$\sigma_2$ which we wish to show are additive.  We can find the
associated channels $N_1'$ and $N_2'$, with the capacity maximized
when the average input density matrix is $\rho_1$ and $\rho_2$, respectively.  
By our hypothesis of additivity of channel capacity, the
tensor product channel $N_1' \otimes N_2'$ has capacity equal to the
sum of the capacities of $N_1'$ and $N_2'$.  If we can now analyze
the capacity of the channel $N_1' \otimes N_2'$ carefully, we might
be able to show that the
entanglement of formation of $E_F(\sigma_1 \otimes \sigma_2)$ is 
indeed the sum of $E_F(\sigma_1)$ and $E_F(\sigma_2)$.  We
do not know how to define such a channel $N'$ satisfying 
(\ref{whatwewant}).  What we actually do
is find a channel whose capacity is close to (\ref{whatwewant}), or 
more precisely a sequence of channels approximating (\ref{whatwewant})
in the asymptotic limit.  
It turns out that this will be adequate to prove the desired theorem.

We now give the definition of our new channel $N'$.   It
takes as its input, the input to 
the channel $N$, along with $k$ additional classical bits (formally, this 
is actually a $2^k$-dimensional Hilbert space on which the first action of 
the channel is to measure it in the canonical basis). 
With probability $q$ the channel
$N'$ sends the first part of its input through the 
channel $N$ and discards the classical bits; with probability
$1-q$ the channel $N$ makes a measurement on the first part of the input, 
and uses the results of this
measurement to decide whether or not to send the auxiliary classical
bits.  When the auxiliary classical bits are not sent, an erasure symbol 
is sent to the receiver instead.  When the auxiliary classical bits are 
sent, they are labeled, so the receiver knows whether he is receiving the 
output of the original channel or the auxiliary bits.

What is the capacity of this new channel $N'$?  Let ${\bf E}$ be
the element of
the POVM measurement in the case that we send the auxiliary
bits 
(so $I-{\bf E}$ is the element of the POVM in the case that we do not send 
these bits).  Now, we claim that for some 
set of vectors $\ket{v_i}$ and some associated set of probabilities 
$p_i$, the optimum signal states of this new channel $N'$ will be 
$\proj{v_i}\otimes\proj{b}$ with associated probabilities $p_i/2^k$, 
where $b$ ranges over all values of the classical bits.\footnote{This
just says that we want to use the classical part of the channel 
as efficiently as possible.  The formal proof is straightforward:
First, we show that it doesn't help to send superpositions of 
the auxiliary bits, so we can assume that the signal states are indeed of
the form $\proj{v_i}\otimes  \proj{b}$.  Next, we show that if two signal
$\proj{v_i} \otimes \proj{b_1}$ and $\proj{v_i} \otimes \proj{b_2}$,
so not have the same probabilities associated with them,  
a greater capacity can be achieved by making these probabilities equal.}

We now can find bounds on the capacity of $N'$.   Let $\ket{v_i}$ and $p_i$ be
the optimal signal states and probabilities for $\chi_{N'}(\rho)$.
We compute
\begin{eqnarray}
\chi_{N'}(\rho) &=& q\left(  
H(N( \sum_i p_i \proj{v_i}) -
\sum_i p_i H(N( \proj{v_i}) 
\right) \nonumber \\
&+& (1-q) k \sum_i p_i \Tr \,  {\bf E} \proj{v_i} \nonumber \\
&+&(1-q)\left( H_2(\Tr \, {\bf E} \sum_i p_i \proj{v_i} ) 
- \sum_i p_i H_2(\Tr \, {\bf E} \proj{v_i}) \right),
\label{firstcapNprime}
\end{eqnarray}
where $H_2$ is the binary entropy function 
$H_2(x) = -x\log x - (1-x) \log (1-x)$.
The first term is the information associated with the channel $N$, 
the second that associated with the auxiliary classical bits, and
the third the information associated with the measurement ${\bf E}$.

Let $\rho = \sum_i p_i \proj{v_i}$ and let $\sigma$ be the associated 
entangled state.   
We can now deduce from (\ref{firstcapNprime}) that
\begin{equation}
\chi_{N'}(\rho) = q \chi_N(\rho) + (1-q) k \Tr \, {\bf E} \rho + (1-q)\delta 
\label{capNprime}
\end{equation}
where $\delta$ is defined as
\[
\delta = H_2(\Tr \, { \bf E} \rho) - \sum_i p_i H_2(\bra{v_i}{\bf E} \ket{v_i}).
\]
Note that $0 \leq \delta \leq 1$, since
$\delta$ is positive by the concavity of 
the entropy function $H_2$, and is at most $1$ since $H_2(p) \leq 1$ for 
$0 \leq p \leq 1$.
Similarly, if we use the optimal states for $\chi_N(\rho)$, we find that
\begin{equation}
\chi_{N'}(\rho) \geq \chi_N(\rho) + (1-q) k \Tr \, {\bf E} \rho 
\label{capNprime2}
\end{equation}

From Eq.~(\ref{capNprime}) and Eq.~(\ref{capNprime2}), we 
find that the $\rho_0$ that maximizes the quantity
\begin{equation}
 q \chi_N(\rho) + (1-q) k \Tr \, {\bf E} \rho ,
\label{capNprimesansdelta}
\end{equation}
we are guaranteed to be
within $1-q$ of the capacity of $N'$.  

We next show that we can find a
measurement ${\bf E}$ such that an arbitrary density matrix $\rho_0$ is
a maximum of (\ref{capNprimesansdelta}).
\begin{lemma}
For any probability $0 < q < 1$, any channel $N$,
and any fixed positive matrix $\rho_0$ over the input space of $N$,
there is a sufficiently large $k_0$ such that for $k \geq k_0$
we can find an $\bf E$ so that the maximum of (\ref{capNprimesansdelta})
occurs at $\rho_0$. (This maximum need not be unique.  If
$\chi_N(\rho)$ is not strictly concave at $\rho_0$, then $\rho_0$ will be
just one of several
points attaining the maximum.)  
\end{lemma}
{\bf Proof:}
It follows from the concavity of von Neumann entropy that 
$\chi_N(\rho)$ is concave in $\rho$.  
The intuition is that we must choose ${\bf E}$
so that the derivative \footnote{This is the intuition.  
This derivative need not actually exist.} 
of (\ref{capNprimesansdelta}) with respect to $\rho$ 
at $\rho_0$ is~0.
Because we only vary over matrices with $\Tr \, \rho = 1$, we can add any
multiple of $I$ to ${\bf E}$ and not change the derivative.
Suppose that in the neighborhood of $\rho_0$, 
\begin{equation}
\chi_N(\rho) \leq \chi_N(\rho_0) + \Tr \, \tau (\rho - \rho_0).
\label{linearrho0}
\end{equation} 
That such an expression exists follows from the concavity of $\chi_N(\rho)$
and the assumption that $\rho_0$ is not on the boundary of the
state space, i.e., has no zero 
eigenvalues.  A full rank 
$\rho_0$ is guaranteed by the MSW correspondence.

To make $\rho_0$ a maximum for Eq.~(\ref{capNprimesansdelta}), we
see from Eq.~(\ref{linearrho0}) that we need to find $\bf E$ so that 
\[
\frac{(1-q)}{q} k {\bf E} =  \lambda I -\tau 
\] 
with $0 \leq {\bf E} \leq I$.  This can be done 
by choosing $k$ and $\lambda$ appropriately. \qed

\pagebreak
Now, suppose we have two entangled states $\sigma_1$ and $\sigma_2$
for which we want to show that
the entanglement of formation is additive.  We create the channels $N_1'$
and $N_2'$ as detailed above.  By the additivity of
channel capacity (which we're assuming), the signal states of the
tensor product channel can be taken to be $\smallket{v^{(1)}_i} \smallket{b_1} 
\otimes \smallket{v^{(2)}_j} \smallket{b_2}$
for $b_1$, $b_2$ any $k$-bit strings, 
with probability $p_i^{(1)}p_j^{(2)}/2^{2k}$.  This gives a 
bound on the channel capacity 
of at most
\begin{eqnarray}
\chi_{N_1' \otimes N_2'} &\leq &
q \left(H({N_1}(\rho_1)) - E_F(\sigma_1)\right) 
+ (1-q) k \Tr \, {\bf E}_1 \rho_1 +
 \nonumber \\
&& + q \left(H({N_2}(\rho_2)) - E_F(\sigma_2) \right)
+ (1-q) k \Tr \, {\bf E}_2 \rho_2
+ 2 (1-q)
\label{tensorcap}
\end{eqnarray}
The $2(1-q)$ term at the end comes from the fact that the formula
(\ref{capNprimesansdelta}) is within $1-q$ of the capacity.
Now, we want to show that we can find a larger capacity than this if 
there is a better decomposition of $\sigma_1 \otimes \sigma_2$, i.e.,
if the entanglement of formation of $\sigma_1 \otimes \sigma_2$ is
not additive.  The 
central idea here is to let $q$ go to 1; this forces $k$ to simultaneously
go to $\infty$.  There is a
contribution from entangled states, which goes as $q^2$, 
a contribution from the auxiliary $k$-bit classical channel, which goes as 
$(1-q) k$, but which is equal in both cases, and a contribution
from unentangled states, which goes as $q(1-q)$.  As $q$ goes to 1, the
contribution from the entangled states dominates the difference.

Suppose there are a set of entangled states which give
a smaller entanglement of formation for
$\sigma_1 \otimes \sigma_2$ than $E_F\sigma_1 + E_F\sigma_2$.  
By the MSW correspondence, this gives a set of signal states
for the map $N_1 \otimes N_2$ which yield a larger constrained capacity
than $\chi_{N_1}(\rho_1) + \chi_{N_2}(\rho_2)$.  We define
this set of signal states for $N_1 \otimes N_2$
to be the states $\proj{\phi_i}$, and let the associated probabilities be
$\pi_i$.  
Now, using the $\ket{\phi_i}$ as
signal states in $N_1' \otimes N_2'$ shows that
\begin{eqnarray}
\chi_{N_1' \otimes N_2'}
&\geq& q^2H(N_1\otimes N_2(\rho_1 \otimes \rho_2)) 
-q^2 E_F(\sigma_1 \otimes \sigma_2)\nonumber\\ &&
 + (1-q) k \Tr \, {\bf E}_2 \rho_2 \nonumber 
\label{entangledcap}
\end{eqnarray}
This estimate comes from considering the information transmitted by
the signal states $\proj{\phi_i}$ in the case (occurring with probability 
$q^2$) when the channels operate as $N_1 \otimes N_2$, as well as the
information transmitted by the $k$ classical bits.

We now consider the difference between this lower bound 
(\ref{entangledcap}) for the capacity of $N'_1 \otimes N'_2$ and the upper 
bound (\ref{tensorcap})
we showed for the capacity using tensor product signal 
states.
In this difference, the terms containing $(1-q)k$ cancel out.  The
remaining terms give
\begin{eqnarray*}
0&\geq& q E_F(\sigma_1) + q E_F (\sigma_2) - q^2 E_F(\sigma_1\otimes\sigma_2)
-2 (1-q) \\
&& - q (1-q) H(N_1(\rho_1)) - q (1-q) H(N_2(\rho_2)).
\end{eqnarray*}
For $q$ sufficiently close to 1, the $(1-q)$ terms can be
made arbitrarily small, and
$q$ and $q^2$ are both arbitrarily close to 1.   This difference can thus 
be made positive if 
the entanglement of formation is strictly subadditive, contradicting our 
assumption that the Holevo channel capacity is additive. 

\newpage
\section{The linear programming formulation}
\label{sec-lp}

We now give the linear programming duality formulation for the
constrained capacity problem.
Recall the definition of the constrained Holevo capacity
\begin{equation}
\chi_N(\rho) = \max_{\{{p_i,\, \ket{\phi_i}\}}\atop {\sum_i p_i 
\proj{\phi_i} = \rho}} H(N( \sum_i p_i \proj{\phi_i})) -
\sum_i p_i H(N( \proj{\phi_i})) 
\label{second-constrained}
\end{equation}
This is a linear program, and as such it has a formulation of a dual
problem that
also gives the maximum value.  This dual problem is crucial to several
of our proofs.  
For this paper, we only deal with channels having finite dimensional
input and output spaces.  For infinite dimensional channels, the duality 
theorem fails unless the maxima are replaced by suprema.  We have not
analyzed the effects this has on the proof of our equivalence theorem, but 
even if it still holds the proofs will become more complicated.

By the duality theorem for linear programming 
there is
another expression for $E_F(\sigma_1)$.  This was observed
in \cite{AB,Shor-compute}.  It is
\begin{equation}
\chi_N(\rho) = H(N(\rho)) - f(\rho)
\label{dualchi}
\end{equation}
where $f$ is the linear function defined by the maximization
\begin{equation}
\max_{f} f(\rho) \mathrm{\ such\ that\ }
f(\proj{v}) \leq H(N(\proj{v})) \quad \mathrm{for\ all\ }\ket{v} \in 
{\cal H}_{\mathrm{in}},
\label{dualprog}
\end{equation}
Here  ${\cal H}_{{\mathrm{in}}}$ is the input space for $N$ 
and the maximum is taken over all linear functions 
\[
f(\rho) = \Tr \, \tau \rho.
\]

Eqs.~(\ref{dualchi}) and (\ref{dualprog}) can be proved if $\rho$ is 
full rank by using 
the duality theorem of linear programming. The duality theorem applies
directly if there are only a finite number of possible signal
states allowed, showing the equality
of the modified version of Eqs.~(\ref{second-constrained}) and
(\ref{dualchi})
where the constraints in (\ref{dualprog}) are limited to a finite number of 
possible signal states $\ket{v_i}$, which are also the only signal states 
allowed in the capacity calculation (\ref{second-constrained}).
To extend from all finite collections of signal states $\proj{v_i}$ to
all $\proj{v}$, we need to show that we can find a compact set of 
linear functions $f(\rho) = \Tr \, \tau \rho$ which suffice to satisfy
Eq.~(\ref{dualprog}).  We can then use compactness to show that
a limit of these functions
exists, where in the limit Eqs.~(\ref{second-constrained}) and
(\ref{dualprog}) must hold on a countable set of possible signal 
states $\ket{v_i}$ dense in the set of unit vectors, thus showing that they
hold on the set of all unit vectors $\ket{v}$.  
The compactness follows from $\rho$ being full rank,
and $H(N(\proj{v})) \leq \log d_\mathrm{out}$ for all $\proj{v}$, where
$d_\mathrm{out}$ is the dimension of the output space of $N$.  
The case where $\rho$ is not full rank can be proved by using
the observation that
the only values of the function $f$ which are relevant in this case
are those in the support of $\rho$.

Equality must hold in (\ref{dualprog})
for those $\ket{v}$ which are signal states in an
optimal decomposition.  This can be seen by considering
the inequalities
\begin{eqnarray*}
\chi_N(\rho) &=& H(N(\rho)) - \sum_i p_i H(N(\proj{v_i}))\\
&\leq& 
H(N(\rho)) - \sum_i p_i f(\proj{v_i})\\
&=&
H(N(\rho)) -  f(\rho)
\end{eqnarray*}
For equality to hold, it must hold in all the terms in the summation,
which are exactly the signal states $\ket{v_i}$. 

\section{Additivity of $E_F$ implies strong superadditivity of $E_F$}
\label{sec-iii-iv}

In this section, we will show that additivity of entanglement of
formation implies strong superadditivity of entanglement of formation.
Another proof was discovered independently by 
Pomeransky~\cite{ssa-from-aef}; it is quite similar, although it
is expressed using different terminology.

We first give the statement of strong superadditivity.
Assume we have a quadripartite density matrix $\sigma$ whose four parts are 
$A1$, $A2$, $B1$ and $B2$.  The statement of strong 
superadditivity is that
\begin{equation}
E_{F} (\sigma) \geq E_{F} (\Tr_2 \sigma) + E_{F} (\Tr_1 \sigma)
\label{strsup-gen}
\end{equation}
where $E_{F}$ is the entanglement of formation when the state
is considered as a bipartite state where the two parts are $A$ and $B$; 
that is,
\begin{equation} 
E_{F} (\sigma) = 
\min_{\{p_i,\,{\ket{\phi_i}\}}\atop {\sum_i p_i \proj{\phi_i} = \sigma}}
\sum_i p_i  H(\Tr_B \proj{\phi_i}).
\end{equation}

First, we show that it is sufficient to prove this when $\sigma$ is a pure 
state.  Consider the optimal decomposition of
$\sigma = \sum_i \pi_i \proj{\phi_i}$.  We can apply the theorem
of strong subadditivity to the pure states $\proj{\phi_i}$ to obtain
decompositions 
$\Tr_1 \proj{\phi_i} = \sum_j p_{i,j}^{(1)} \proj{v^{(1)}_{i,j}}$ and 
$\Tr_2 \proj{\phi_i} = \sum_{j} p^{(2)}_{i,j} \proj{v^{(2)}_{i,j}}$ 
so that
\[
H(\Tr_B \proj{\phi_i}) \geq
\sum_j p_{i,j}^{(1)} H(\Tr_B \proj{v_{i,j}^{(1)}})
+ \sum_j p_{i,j}^{(2)} H(\Tr_B \proj{v_{i,j}^{(2)}} ).
\]
Summing these inequalities over $i$ gives the desired inequality.

We now show that additivity of $E_F$ implies
strong superadditivity of $E_F$.  Let $\ket{\phi}$ be a quadripartite
pure state for we wish to show strong superadditivity.
We define $\sigma_1 = \Tr_2\proj{\phi}$ and 
$\sigma_2 = \Tr_1\proj{\phi}$.  Now, let us use the MSW correspondence
to find channels $N_1$ and $N_2$ and density matrices
$\rho_1$ and $\rho_2$ such that 
\[
N_1 (\rho_1) = \Tr_B\sigma_1 \quad \mathrm{and} \quad N_2 (\rho_2) = 
\Tr_B \sigma_2 
\]
and
\begin{eqnarray*}
\chi_{N_1}(\rho_1) &=& H(N_1(\rho_1)) - E_F(\sigma_1)\\
\chi_{N_2}(\rho_2) &=& H(N_2(\rho_2)) - E_F(\sigma_2)
\end{eqnarray*}

We first do an easy case which illustrates how the proof works without 
introducing additional complexities.  Let
$d_1$ and $d_2$ be the dimensions of the input spaces of $N_1$ and $N_2$.
In the easy case, we
assume that there are $d_1^2$ 
linearly independent signal states in an optimal decomposition of
$\rho_1$ for $\chi_{N_1}(\rho_1)$, and $d_2^2$ linearly independent 
signal states in an optimal decomposition of $\rho_2$ 
for $\chi_{N_2}(\rho_2)$.  Let these sets of signal states
be $\proj{v_i^{(1)}}$ with probabilities $p_i^{(1)}$, and 
$\proj{v_j^{(2)}}$ with probabilities $p_j^{(2)}$, respectively.
It now follows from our assumption of the additivity of entanglement
of formation that an optimal ensemble of signal states for 
$\chi_{N_1 \otimes N_2}(\rho_1 \otimes \rho_2)$ is
$\smallket{v_i^{(1)}} \otimes \smallket{v_j^{(2)}}$ with probability
$p_i^{(1)} p_j^{(2)}$.  

Now, let us consider the dual linear function $f_T$ for the tensor
product channel
$N_1 \otimes N_2$.  Since we assumed that 
entanglement of formation is additive, 
by the MSW correspondence $\chi_N(\rho)$ is also additive.  
We claim that the dual function $f_T$ must satisfy
\begin{equation}
f_T(\proj{v^{(1)}_i} \otimes \proj{v^{(2)}_j})
= H(N_1(\proj{v_i^{(1)}})) + H(N_2(\proj{v_j^{(2)}}))
\end{equation}
for all signal states $\smallket{v^{(1)}_i}\smallket{v^{(2)}_j}$.
This is simply because equality must hold in the inequality (\ref{dualprog})
for all signal states.
However, we now have that $f_T$ is 
a linear function 
in a $d_1^2 d_2^2 -1 $ dimensional space
which has been specified on 
$d_1^2 d_2^2$ linearly
independent points; this implies 
that the linear function $f_T$ is uniquely defined.  It is easy to see that it
thus must be the case that
\begin{equation}
f_T(\rho)  = f_1 (\Tr_2 \rho) + f_2 (\Tr_1 \rho),
\label{addf}
\end{equation} 
as this holds for the $d_1^2 d_2^2$ signal states
We now let $\proj{\psi}$ be the preimage of $\Tr_B \proj{\phi}$ under 
the channel $N_1 \otimes N_2$.  
We have, from the equations (\ref{dualprog}) and (\ref{addf}), that
\begin{equation}
f_1 (\Tr_2 \proj{\psi}) + f_2(\Tr_1 \proj{\psi}) \leq 
H(N_1 \otimes N_2 ( \proj{\psi} ) ).
\label{pre-ssaEF}
\end{equation}
But recall that 
\begin{eqnarray}
f_1 (\Tr_2 \proj{\psi}) &=& E_F(\sigma_1), \nonumber \\
f_2 (\Tr_1 \proj{\psi}) &=& E_F(\sigma_2),
\label{fisEF}
\end{eqnarray}
because (\ref{dualprog}) holds with equality for signal states, 
and that
\[
N_1 \otimes N_2 (\proj{\psi}) = \Tr_B \proj{\phi}.
\]
Thus, substituting into (\ref{pre-ssaEF}), we find that
\[
E_F( \sigma_1) + E_F(\sigma_2) \leq H(\Tr_B \proj{\phi}),
\]
which is the statement for the strong superadditivity of entanglement 
of formation of the pure state $\proj{\phi}$.

We now consider the case where there are fewer than $d_i^2$ signal states
for $\chi_{N_i}(\rho_i)$, $i=1,2$.  We still
know that the average density matrices of the signal states
for $N_1$ and $N_2$ are $\rho_1$ and $\rho_2$, and that the
support of these two matrices are the entire input 
spaces ${\cal H}_{1,\mathrm{in}}$ and 
${\cal H}_{2,\mathrm{in}}$.  The argument will go as 
before if we can again show that the dual function $f_T$ must be
$f_1(\Tr_2 \rho) + f_2(\Tr_1 \rho)$.  In this case we
do not know $d_1^2d_2^2$ points of the function $f_T$, and thus cannot 
use the same argument as above to show that $f_T$ is determined.  
However, there is more information that we have available.
Namely, we know that in the neighborhood of the signal states 
$\smallket{v_i^{(1)}}$,
the entropy $H(N_1(\proj{v}))$ must be at least the dual 
function $f_1 = \Tr \, \tau_1 \proj{v}$, and that these two functions
are equal at the signal states.  If we assume that the derivative
of $H(N_1(\proj{v}))$ exists at $\proj{v_i^{(1)}}$, then we can conclude 
that this is also the derivative of $f_1 = \Tr \, \tau_1 \proj{v}$.  
For the time being we will assume that the first derivative of
this entropy function does in fact exist.\footnote{In fact, I believe the
function is smooth enough that these derivatives do exist. 
However, we find it easier to deal
with the cases where $N_1(\proj{v})$ has zero eigenvalues by expressing
$N_1$ and $N_2$ as a limit of nonsingular completely positive maps.} 

We need a lemma.
\begin{lemma}
\label{spanlemma}
Suppose that we have a set of unit vectors $\ket{v_i}$ that span a
Hilbert space ${\cal H}$.
If we are given the value of $f$ at all the vectors $\ket{v_i}$ as well
as the value of the first derivative of $f$,
\[
\lim_{\epsilon \rightarrow 0} \frac{1}{\epsilon} \Big(
f(\proj{v_i}) - 
f\big((\sqrt{(1-\epsilon^2}\ket{v_i}+ \epsilon \ket{w_i})
(\sqrt{1-\epsilon^2}\bra{v_i}+ \epsilon \bra{w_i})\big)\Big)
\]
at all the vectors $\ket{v_i}$ and for all orthogonal $\ket{w}$, 
then $f$ is completely determined.  
\end{lemma}
{\bf Proof:}
Let us use the representation $f(\rho) = \Tr \, \tau \rho$
(we do not need a constant term on the right hand side
because we need only specify $f$ on
trace 1 matrices).  Suppose that $\braket{v_i }{w} = 0$.  We
compute the derivative at $\smallket{v_i}$ in the $\ket{w}$ direction:
\begin{eqnarray}
\left(\sqrt{1-\epsilon^2} \smallbra{v_i} + \epsilon \smallbra{w}\right)\,
\tau\,
\left(\sqrt{1-\epsilon^2} \smallket{v_i} + \epsilon \smallket{w}\right)
- \smallbra{v_i} \tau \smallket{v_i} \quad \quad && \nonumber \\ 
\approx
\epsilon \left(\smallbra{v_i} \,\tau\, \smallket{w} + 
\smallbra{w} \,\tau\, \smallket{v_i}\right). &&
\label{wdir}
\end{eqnarray}
The derivative in the $i \ket{w}$ direction gives
\begin{equation}
i \left(\smallbra{v_i} \,\tau\,  \smallket{w} - \smallbra{w} 
\,\tau\, \smallket{v_i}\right),
\label{iwdir}
\end{equation}
so a linear combination of (\ref{wdir}) and (\ref{iwdir}) shows
that the value of 
$\smallbra{v_i} \,\tau\, \smallket{w}$ is determined for all $\ket{w}$
orthogonal to $\smallket{v_i}$.
We also know the value of 
\[
\smallbra{v_i} \,\tau\, \smallket{v_i},
\]
it follows that the value of 
\[
\smallbra{v_i} \tau \ket{w}
\]
is determined for all $\ket{w}$.
Since the $\smallbra{v_i}$ span the vector space, this determines 
the value of
\[
\bra{u} \tau \ket{w}
\]
for all $\bra{u}$ and all $\ket{w}$, thus determining the matrix $\tau$.
\qed

We now need to compute the derivative of the entropy of $N_1$.  Let
\[
N_1(\rho) = \sum_i A_i \rho A_i^\dag
\]
with $\sum_i A_i^\dag A_i = I$.
Then if $\Tr \, \sigma = 0$, 
\begin{eqnarray}
\nonumber
H(N_1(\rho + \epsilon \sigma)) - H(N_1(\rho)) &\approx&
-\epsilon \Tr \left[ (I  + \log(N(\rho)) N_1(\sigma)\right] \\
&=& - \epsilon \Tr \left( 
\sigma \sum_k A^\dag_k \big(\log N_1(\rho)\big) A_k \right)
\end{eqnarray}
Now, if the entanglement of formation is additive, then the derivative
of $H(N_1 \otimes N_2)$ at the
tensor product signal states $\proj{v_i^{(1)}} \otimes \proj{v_j^{(2)}}$
must also match the derivative of the function $f_T$ at these points.  
We calculate:
\begin{eqnarray*}
&&H(N_1\otimes N_2(\rho + \epsilon \sigma)) - H(N_1\otimes N_2(\rho)) \quad
\\
&&\quad \quad \approx
- \epsilon 
\Tr 
\left(\sigma \sum_{k_1,k_2} (A^{(1)\dag}_{k_1}\otimes A^{(2)\dag}_{k_2}) 
(\log (N_1\otimes N_2(\rho)))
 ( A^{(1)}_{k_1} \otimes A^{(2)}_{k_2})\right).
\end{eqnarray*}
Now at a point $\rho = \rho_1 \otimes \rho_2$, 
\begin{eqnarray*}
&&\sum_{k_1,k_2} (A^{(1)\dag}_{k_1}\otimes A^{(2)\dag}_{k_2}) 
(\log N_1\otimes N_2(\rho)) ( A^{(1)}_{k_1} \otimes A^{(2)}_{k_2})\\
&&\quad\quad= 
\big(\sum_{k_1} A^{(1)\dag}_{k_1} \log N_1(\rho_1)  A^{(1)}_{k_1}\big)  
\otimes I
+
I \otimes 
\big(\sum_{k_2} (A^{(2)\dag}_{k_2} \log N_2(\rho_2)  A^{(2)}_{k_2}\big)  ,
\end{eqnarray*}
showing that at the states 
$\smallket{v_i^{(1)}} \otimes \smallket{v_j^{(2)}}$, we
have not only that $f_T = f_1 + f_2$, but that the first 
derivatives (for directions $\sigma$ with $\Tr \, \sigma = 0$)
are equal as well.  Since the states
$\smallket{v_i^{(1)}} \otimes \smallket{v_j^{(2)}}$ span the vector space,
Lemma~\ref{spanlemma} shows
that  $f_T = f_1 + f_2$ everywhere, giving us the last element of 
the proof.

The one thing remaining to do to show that the assumption that the
first derivative 
of entropy exists everywhere is unnecessary.  
It suffices to show that there are dual functions $f_T = f_1 + f_2$
such that Eq.~(\ref{pre-ssaEF}) holds.  We do this by taking limits.  
For $x = 1,2$ let
$N^{(q)}_x$ be the quantum channel
\[
N^{(q)}_x(\rho) = N_x(\rho) + (1-q) \frac{1}{d_{\mathrm{out},x}} I
\]
which averages the map $N_x$ with the maximally mixed 
state ${I}/{d_{\mathrm{out},x}}$.  Let 
$N_T^{(q)} = N_1^{(q)} \otimes N_2^{(q)}$.
We need to show that some limits of the dual functions $f_1^{(q)}$,
$f_2^{(q)}$ and $f_T^{(q)}$ exist.  By continuity of $N^{(q)}_x$, they
will be forced to have the desired properties 
(\ref{addf}), (\ref{pre-ssaEF}), and (\ref{fisEF}).
Let $\rho_T = \rho_1 \otimes \rho_2$.
Now, 
$f_T^{(q)}$ is a linear function with $f_T^{(q)}(\rho_T) \geq 0$ and 
$f_T^{(q)}(\rho) \leq \log d_{\mathrm{out},T}$ for all $\rho$, so the 
$f_T^{(q)}$ lie in 
a compact set.  Thus, some subsequence of $f_T^{(q)}$ has a limit as
$q \rightarrow 1$.  The same argument applies to $f_1^{(q)}$ and $f_2^{(q)}$,
so by taking these limits we find that the functions $f_x^{(1)}$ have the 
desired properties, completing our proof.

\section{Additivity of $\min H(N)$ implies additivity of $E_F$.}
\label{sec-i-iii}

Suppose that we have two bipartite states for which we wish to prove that
the entanglement of formation is additive.  
We use the MSW correspondence to convert this problem to a question
about the Holevo capacity with a constrained average signal state.  We thus
now have two quantum channels $N_1$ and $N_2$, and two states $\rho_1$ and
$\rho_2$.  
We want to show that 
\[
\chi_{N_1\otimes N_2}(\rho_1 \otimes \rho_2) =
\chi_{N_1}(\rho_1) + \chi_{N_2}(\rho_2).
\]
In fact, we need only prove the $\leq$ direction of the inequality,
as the $\geq$ direction is easy.

Let $\smallket{v_i^{(1)}}$ and $\smallket{v_i^{(2)}}$ be optimal
sets of signal states for $\chi_{N_1}(\rho_1)$ and $\chi_{N_2}(\rho_2)$, so that
\[
\chi_{N_1}(\rho_1) = H(N_1(\rho_1)) - \sum_i p_i^{(1)} N(\proj{v_i^{(1)}})
\]
where $\rho_1 = \sum_i p_i^{(1)} \proj{v_i^{(1)}}$, and similarly
for $N_2$.  By the linear programming dual formulation in 
Section~\ref{sec-lp}, we have that there is a matrix $\tau_1$ 
such that
\[
\chi_{N_1}(\rho_1) = H(N_1(\rho_1)) - \Tr \, \tau_1 \rho_1
\]
and
\[ 
\Tr \, \tau_1 \rho \leq H(N_1(\rho)
\]
for all $\rho$, with
equality for signal states $\rho = \proj{v_i^{(1)}}$,
and similarly for $\tau_2$ and $N_2$.  
Suppose we could find a channel
$N'_1$ and $N_2'$ such that
\begin{equation}
H(N'_1(\proj{v})) =
H(N_1(\proj{v})) + C_1 - \bra{v} \tau \ket{v} 
\label{wantNprime}
\end{equation}
for all vectors $\ket{v}$ (similarly for $N_2$).  
We know from the linear programming duality theorem that
\begin{eqnarray*}
H(N'_1(\rho)) &=&
H(N_1(\rho)) + C_1 - \Tr \, \tau_1 \rho \\
&\geq& C_1
\end{eqnarray*}
for all input states $\rho$, with equality holding for the signal
states $\rho = \proj{v_i^{(1)}}$.  Thus, the minimum entropy output 
of $N_1'$ is $C_1$ and of $N_2'$ is $C_2$.  
Also, 
\begin{eqnarray*}
\chi_{N_1'}(\rho_1) &=& H(N_1'(\rho_1))) - \sum_i p^{(1)}_i 
H(N_1'(\proj{v^{(1)}_i}))\\
&=& H(N_1'(\rho_1))) - C_1,
\end{eqnarray*}
and similarly for $N_2'$.
Now, if we assume the additivity of minimum entropy, we know that the 
minimum entropy output of $N_1' \otimes N_2'$ has entropy $C_1 + C_2$.  We have
for some probability distribution $\pi_i$ on signal states $\ket{\phi_i}$, that
\begin{eqnarray*}
\chi_{N_1' \otimes N'_2} (\rho_1 \otimes \rho_2) 
&=& H(N_1'\otimes N_2'(\rho_1 \otimes \rho_2))
- \sum_i \pi_i H(N'_1 \otimes N'_2 (\proj{\phi_i})) \\
&\leq& 
H(N_1'(\rho_1)) + H(N_2'(\rho_2)) - C_1 - C_2 \\
&=& \chi_{N_1'}(\rho_1) + \chi_{N_2'}(\rho_2)
\end{eqnarray*}
Now, if we can examine the construction of the channels $N_1'$ and $N_2'$
and show that the additivity of the constrained Holevo capacity for
$N_1'$ and  $N_2'$ implies the additivity of the constrained Holevo capacity
for $N_1$ and $N_2$, we will be done.  

We will not be able to 
achieve Eq.~(\ref{wantNprime}) exactly, but will be able to achieve
this approximately, in much the same way we defined $N'$ in 
Section~\ref{sec-ii-iii}.  

Given a channel $N$, 
we define a new channel $N'$.   On input $\rho$, with probability $q$ the
channel $N'$ outputs $N(\rho)$.  With probability $1-q$ the channel
makes a POVM measurement 
with elements ${\bf E}$ and $I-{\bf E}$.  If the measurement outcome
is ${\bf E}$, $N'$ outputs the tensor product of
a pure state signifying that the result was ${\bf E}$ 
and the maximally mixed state on $k$ qubits.  If the result is 
$I-{\bf E}$ the channel $N'$ outputs only a pure state signifying this fact.
We have 
\[
H(N'(\rho)) = qH(N(\rho)) + H_2(q) + (1-q) k \Tr \, {\bf E} \rho +
(1-q) H_2(\Tr \,{\bf E} \rho).
\]
If we choose $k$ and ${\bf E}$ such that 
\[
\frac{(1-q)}{q} k {\bf E} = \lambda I -\tau, 
\]
we will have
\[
H(N'(\proj{v})) = q H(N(\proj{v})) - q \bra{v} \tau \ket{v}
+q\lambda + H_2(q) + (1-q) H_2(\bra{v}{\bf E} \ket{v}).
\]
The minimum entropy $H(N'(\proj{v}))$ is thus at least 
$q \lambda + H_2(q)$.
For signal states $\ket{v_i}$ of $N$, $H(N'(\proj{v_i}))$ is 
at least $q\lambda + H_2(q)$
and at most $q\lambda + H_2(q) + 1-q$.  As $q$ goes to 0, this is
approximately a constant.  We thus see that
\begin{equation}
H(N_1'(\rho_1)) - q \lambda_1 - H_2(q) - (1-q) \leq
\chi_{N'_1}(\rho_1) \leq
H(N_1'(\rho_1)) - q \lambda_1 -H_2(q)
\label{boundsonNprime}
\end{equation}

Now, given two channels $N_1$ and $N_2$, we can prepare $N_1'$ and $N_2'$
as above.  If we assume the additivity of minimum entropy, this implies
the constrained channel capacity satisfies, for the 
optimal input ensembles  $\ket{\phi_i}$, $\pi_i$,
\begin{eqnarray*}
\chi_{N'_1 \otimes N'_2}(\rho_1 \otimes \rho_2) 
&=&
H(N_1'(\rho_1)) + H(N'_2(\rho_2)) 
- \sum_{i} \pi_i H(N'_1\otimes N'_2
(\proj{\phi_i}))\\
&\leq&
H(N_1'(\rho_1)) + H(N'_2(\rho_2)) - q\lambda_1 -q\lambda_2 -2H_2(q)\\
&\leq&
\chi_{N_1'}(\rho_1) + \chi_{N'_2}(\rho_2) + 2(1-q)
\end{eqnarray*}
where the first inequality follows from the assumption of additivity of
the minimum entropy output, and the second from Eq.~(\ref{boundsonNprime}).

We now need to relate $\chi_{N_1'}(\rho_1)$ and $\chi_{N_1}(\rho_1)$.  Suppose
we have an ensemble of signal states $\proj{v_i}$ with 
associated probabilities $p_i$, and such that 
$\sum_i p_i \proj{v_i} = \rho$.  Define $C_{N_1}$ ($C_{N'_1}$) to be
the information transmitted by channel $N_1$ ($N_1'$) using these signal 
states.  We then have
\[
C_{N_1'} = qC_{N_1}  + (1-q) \delta_1
\]
where
\[
\delta_1 = H_2(\Tr \, { \bf E} \rho) - \sum_i p_i H_2(\bra{v_i}{\bf E} \ket{v_i}).
\]
This shows that
\begin{eqnarray*}
q \chi_{N_1}(\rho_1) 
\leq \chi_{N_1'}(\rho_1) \leq 
q \chi_{N_1}(\rho_1) +(1-q)
\end{eqnarray*}
Also, by using the optimal set of signal states for 
$\chi_{N_1\otimes N_2}(\rho_1\otimes \rho_2)$ as signal states for the
channel $N_1'\otimes N_2'$, we find that
\[
\chi_{N'_1 \otimes N'_2}(\rho_1\otimes \rho_2) \geq
q^2\chi_{N_1 \otimes N_2}(\rho_1\otimes \rho_2)
\]
since with probability $q^2$, the channel $N_1' \otimes N_2'$ simulates
$N_1 \otimes N_2$.  
Thus, we have that
\begin{eqnarray*}
\chi_{N_1\otimes N_2}(\rho_1\otimes \rho_2) &\leq& 
q^{-2} \chi_{N'_1\otimes N'_2} (\rho_1\otimes \rho_2)\\
&\leq& q^{-2} (\chi_{N'_1}(\rho_1) + \chi_{N'_2}(\rho_2)) + 2(1-q)q^{-2}\\
&\leq& q^{-1} (\chi_{N_1}(\rho_1) + \chi_{N_2}(\rho_2)) + 4(1-q)q^{-2}
\end{eqnarray*}
holds for all $q$, $0<q<1$.  Letting $q$ go to 1, we have subadditivity
of the constrained Holevo capacity, implying additivity of
the entanglement of formation.

\section{Implications of strong superadditivity of $E_F$.}
\label{sec-iv-all}

All three additivity properties (i) to (iii) follow easily from the assumption
of strong superadditivity of $E_F$.  The additivity of $E_F$ follows
trivially from this assumption.  That the additivity of $\chi_N$ follows
is known \cite{MSW}.  We repeat this argument below for completeness.
Recall the definition of $\chi_N$:  
\begin{equation}
\chi_N = \max_{\{p_i,\, \ket{\phi_i}\} }
H(N( \sum_i p_i \proj{\phi_i})) -
\sum_i p_i H(N( \proj{\phi_i}))
\end{equation}
Suppose that this the maximum is attained at an ensemble $p_i, \ket{\phi_i}$
that is not a tensor product distribution.  If we replace this ensemble with
the product of the marginal ensembles, the concavity of von Neumann entropy
implies that the first term increases, and the superadditivity of entanglement
of formation implies that the second term decreases, showing that we can
do at least as well by using a tensor product distribution, and that
$\chi_N$ is thus additive.

Finally, the proof that strong superadditivity of $E_F$ implies
additivity of minimum output entropy is equally easy, although I
am not aware of its being in the literature. 
Suppose that we have a minimum entropy output
$\chi_{N_1\otimes N_2}(\proj{\phi})$.  The strong superadditivity of 
$E_F$ implies that there
are ensembles $p_i^{(1)}, \smallket{v_i^{(1)}}$ and 
$p_i^{(2)}, \smallket{v_i^{(2)}}$
such that
\[
H({N_1\otimes N_2}(\proj{\phi})) \geq 
\sum_i p_i^{(1)} H( N_1(\proj{v_i^{(1)}})
+ \sum_i p_i^{(2)} H( N_2(\proj{v_i^{(2)}}).
\]
But the two sums on the right hand side are averages, so there must
be one quantum state in each of these sums have smaller output entropy
than the average output entropy; this shows additivity of the minimum entropy
output. 

\section{Additivity of $\chi_N$ or of $E_F$ implies additivity of 
$\min H(N)$.}
\label{sec-all-i}

Suppose we have two channels $N_1$ and $N_2$ which map their input
onto $d$-dimensional output spaces.  We can assume that the two output
dimensions are the same by embedding the smaller dimensional output space
into a larger dimensional one.\footnote{This is not necessary
for the proof, but 
it reduces the number of subscripts required to express it.}
We will define two new channels $N_1'$ and $N_2'$.  
The channel $N_1'$ 
will take as input the tensor product of the input space of channel $N_1$ 
and an integer between $0$ and $d^2-1$.  
Now, let $X_0$ $\dots$ $X_{d^2-1}$ be the 
$d$-dimensional generalization of the Pauli matrices:
$X_{da +b} = T^a R^b$, where $T$ takes $\ket{j}$ to 
$\ket{j+1 (\mathrm{mod}\ d)}$ and $R$ takes $\ket{j}$ to $e^{2\pi i j/d} 
\ket{j}$. Let
\[
N_1'(\rho \otimes \proj{i}) = X_i N_1(\rho) X^{\dag}_i.
\]

Now, suppose that $\proj{v_1}$ is the input giving the minimal entropy output
$N_1(\proj{v_1}$).  We claim that a good ensemble of 
signal states for the channel $N_1'$
is $\proj{v_1} \otimes \proj{i}$, where $i = 0,1, \ldots, d^2-1$, with 
equal probabilities.  This is because for this set of 
signal states, the first term in the formula for Holevo 
capacity~(\ref{def-chi}) is 
maximized (taking any state $\rho$ and averaging over all 
$X_i \rho X^{\dag}_i$ gives the maximally mixed state, which has
the largest possible entropy in $d$ dimensions), 
and the second term is minimized.  The same holds for the channel $N_2'$.  
Now, suppose there is some state $\proj{w}$ which has smaller output 
entropy for the channel
$N_1 \otimes N_2$ than $H( N_1(\proj{v_1}) + H( N_2(\proj{v_2}) )$. 
We can use the ensemble containing 
states $\proj{w} \otimes \proj{i_1, i_2}$, for 
$i_1, i_2$ $=$ $0$ $\ldots$ $d^2-1$, 
with equal probabilities,
to obtain a larger capacity for the 
tensor product channel $N_1' \otimes N_2'$.

The above argument works equally well to show that additivity
of entanglement of formation implies additivity of minimum entropy output.
We know that to achieve the maximum capacity, the average output 
state must be the maximally mixed state, so we can equally well use the
fact that the constrained Holevo capacity $\chi_N(\rho)$ is additive to 
show that the minimum entropy output is additive.

\section{Discussion}
\label{sec-discuss}

We have shown that four open additivity questions are equivalent.  This
makes these questions of even greater interest to quantum information
theorists.  Unfortunately, our techniques do not appear to be powerful
enough to resolve these questions.  

The relative difficulty of the proofs of the implications given in this paper
would seem to imply
that of these equivalent conjectures, additivity of minimum entropy 
output is in some sense the ``easiest'' and strong superadditivity
of $E_F$ is in some sense the ``hardest.''  One might thus try to prove 
additivity of the minimum entropy output as a means of solving all of 
these equivalent conjectures.  One step towards solving this problem
might be a proof that the tensor product of states producing locally 
minimum output entropy gives a local minimum of output entropy in the
tensor product channel.

\subsection*{Acknowledgments}
I would like to thank Beth Ruskai for calling my attention to the
papers \cite{AB,MSW} and for helpful discussions, and to 
Beth Ruskai, Keiji Matsumoto, and an anonymous referee for useful 
comments on drafts of this paper.

\end{document}